# Gossiping with Multiple Messages

Sujay Sanghavi, Bruce Hajek, and Laurent Massoulie
sanghavi@mit.edu, b-hajek@uiuc.edu, laurent.massoulie@thomson.net

July 10, 2018

**Abstract**

This paper investigates the dissemination of multiple pieces of information in large networks where users contact each other in a random uncoordinated manner, and users upload one piece per unit time. The underlying motivation is the design and analysis of piece selection protocols for peer-to-peer networks which disseminate files by dividing them into pieces. We first investigate one-sided protocols, where piece selection is based on the states of either the transmitter or the receiver. We show that any such protocol relying only on pushes, or alternatively only on pulls, is inefficient in disseminating all pieces to all users. We propose a hybrid one-sided piece selection protocol – INTERLEAVE – and show that by using both pushes and pulls it disseminates $k$ pieces from a single source to $n$ users in $10(k + \log n)$ time, while obeying the constraint that each user can upload at most one piece in one unit of time, with high probability for large $n$. An optimal, unrealistic centralized protocol would take $k + \log_2 n$ time in this setting. Moreover, efficient dissemination is also possible if the source implements forward erasure coding, and users push the latest-released coded pieces (but do not pull). We also investigate two-sided protocols where piece selection is based on the states of both the transmitter and the receiver. We show that it is possible to disseminate $n$ pieces to $n$ users in $n + O(\log n)$ time, starting from an initial state where each user has a unique piece.

## 1 Introduction

Peer-to-peer systems are decentralized networks enabling users to contribute resources for mutual benefit. One of the main applications of such networks is the cost-effective distribution of bandwidth-intensive content from one source, or a few sources, to many users simultaneously. Peer-to-peer networks such as eDonkey and BitTorrent, which routinely serve files hundreds of megabytes in length to thousands of users, now account for a sizable share of all Internet traffic [1]. Examples of content distribution systems leveraging end-user resources are [2–5]. The service capacity in such systems can grow with the number of users, making them scalable and efficient in servicing a large number of users [6, 7].

File dissemination networks can be broadly categorized into *structured* and *unstructured* networks. Structured networks such as [3–5] rely on a specific structured pattern of interconnections among users to deliver the advantages of scalability and robustness. The structured pattern is first set up in a decentralized fashion, and data is then disseminated using this infrastructure. Unstructured networks have minimal infrastructure, and instead rely on randomization to provide load balancing, robustness, and scalability. For example, in the BitTorrent [2] system the only available infrastructure is a tracker of the addresses of users interested in obtaining the file. Each user acquires a random list of other users from the tracker, who become neighbors. Each user's actions are based on local information obtained from its neighbors.

This paper investigates data dissemination in unstructured networks. Initial unstructured approaches [8, 9] advocated uploading the whole file at one go. Users receiving the complete file would then upload it to other users chosen at random. These protocols were motivated in part by earlier theoretical work on random gossip models [10, 11] and epidemics [12]. However, for large files, making users wait to receive the entire file before they can start serving it becomes untenable for two reasons: *(a)* file transfer may take a long time, and during this time the upload capacity of downloading users is wasted, and *(b)* users who have received the file may depart before uploading a complete copy, resulting in the complete file being lost to others. Modern peer-to-peer file dissemination protocols such as BitTorrent take the following alternate approach to speed up dissemination: the file is divided into pieces, and users can start serving individual pieces once they are received, instead of waiting to obtain the entire file.

When the file to be disseminated is divided into multiple pieces, each user has to carry out the task of *piece selection:* deciding which particular piece of the file is to be communicated at any given time, based on local information[1]. These local decisions have a significant impact on the global effectiveness of file dissemination, as the spread of one piece interacts with the spread of other pieces. The motivations of this paper are *(1)* to gain a quantitative analytical understanding of how splitting a file speeds up its dissemination in networks with random user contacts, and *(2)* to understand how the users' local piece selection decisions impact the dissemination of multiple pieces.

In Section 2, we develop a simple model of a peer-to-peer system relaying multiple pieces. We also explain how it captures the speedup obtained from file splitting, and enables us to compare the efficiency of various piece selection protocols. The user contact model is the same as in the classical random gossip process literature [8–11]. In Section 3 we state our main results, and discuss their implications. Sections 4 and 5 contain the lemmas and proofs of the main theorems. Section 6 contains some simulation results on protocol performance when some of the assumptions made in the model are relaxed.

---

[1] In BitTorrent for example, this decision is made by each user based only on the information of its neighbours. Each peer polls its neighbors for their piece collections, and then downloads the *locally rarest* piece, i.e. the piece that has the lowest representation in the peer's neighbors.

# 2   Framework

We now present our peer-to-peer system model. A real-world deployed peer-to-peer network such as BitTorrent is an immensely complex system to model and analyze exactly, and we simplify some aspects for the purposes of tractability. In the following we first describe our framework, and then discuss the assumptions and approximations made.

Consider a network with $n$ users, each of whom wants to receive an entire copy of a file. The file is divided into pieces. All users have the same upload bandwidth, and time is measured in slots. The length of each slot is the time one user takes to upload one piece. Any user receiving a piece in some slot $t$ can upload that piece to other users beginning in slot $t+1$. Thus the spread of one piece interacts with the spread of other pieces via the bandwidth constraint.

Throughout this paper, it is assumed that the users contact each other in the following manner: in each time slot each user chooses a *target*, which is another user chosen uniformly at random from the entire network, independently of any state, history, or other users' choices. Communication in that time slot occurs only between each user and its target. This is the contact and communication model used in the classical single-piece random gossip literature [8–11]. We provide bounds and performance guarantees that hold with high probability for large $n$. Our work goes beyond the classical random gossip literature in that it investigates the simultaneous spread of multiple pieces.

Once a target is chosen, the user undertakes one of the following two possible actions

- *pull:* the user selects a piece it does not currently possess and requests it from the target.
- *push:* the user selects a piece it possesses, and transmits it to the target.

For either of the two actions above, the user needs to make a piece selection. This piece selection is said to be *one-sided* if it is based only on the user's own current state, and not on the state of the target. The piece selection is said to be *two-sided* if it is based on the current states of both the user and the target. In either case the selection is independent of system history or the states of other users. Different ways of making this choice correspond to different protocols. In this paper we evaluate the performance of the protocols as measured by the *completion time*, which is the first time slot that all users have all pieces.

Users have limited upload bandwidth. In this paper this is represented by either a *hard constraint* or a *soft constraint*. In the former, each user can upload at most one piece in any instant of time, while in the latter a user is allowed to upload (potentially) any number of pieces simultaneously. The fact that targets are chosen uniformly at random means that a network with $n$ users most likely has a maximum loading of at most $\log n$ for the case of soft constraints. Soft constraints have previously been analyzed in the random gossip literature, see e.g. [9, 13]. In our work we do not impose any constraints on the download bandwidth of the users. However, due to random user contacts and upload constraints, the average usage of download bandwidth is still one piece per slot.

The following simple calculation, similar to the one in [6], demonstrates the potential speedup that can be had from file splitting. Consider an initial condition where the file is divided into $k$ pieces, and all are initially present at one user, called the source. It is easy to see that the completion time would be at least $k + \log_2 n$ time slots, because it takes at least $k$ slots for the last piece to emerge from the source, and a further $\log_2 n$ slots for that piece to reach all users. This $k + \log_2 n$ lower bound holds even for systems that employ coding. It has been shown in [14], under the same upload and download constraints as in this paper, that a fully centralized scheme can achieve this bound for all $n$ and $k$. Other related work, including [15, 16], with slightly different communication constraints, also points to optimal dissemination times that are close to $k + \log_2 n$. If, however, the file is not divided into pieces, each user can upload data only after receiving a copy of the entire file, which takes $k$ time slots. In this case, complete file dissemination takes at least $k \log_2 n$ time slots, because the number of users having the file can at most double every $k$ time slots. If we denote the ratio of the dissemination time for an unsplit file to the dissemination time of a file split into $k$ pieces as the speedup achieved by splitting the file into $k$ pieces, then

$$\text{optimal splitting speedup} \ = \ \frac{k \log_2 n}{k + \log_2 n}$$

From the above we see that if a decentralized protocol with random user contacts has a completion time close to $k + \log n$ then its performance is close to that of a centralized optimal protocol, while if its completion time is closer to $k \log n$ then it is performing badly, providing little speedup from file splitting.

For large networks, splitting a file into a large number of pieces gives significant speedup gains, but at the expense of increased overhead. For example, two-sided protocols may require users to maintain the current states of their neighbors. This may be hard when there are a large number of small pieces. This is the motivation for investigating piece-selection protocols relying on less than complete information, of which one-sided protocols represent an extreme case. Other overheads arise from network and system considerations, such as the choice of the underlying transport control protocol.

We now briefly discuss the modeling approximations made. Network effects including delays, packet losses due to congestion, and user heterogeneity have been abstracted away: we assume that communicating a piece always takes the same amount of time, between any two users. Also, real-world systems are typically open, with users joining and leaving, while our analysis assumes the simultaneous arrival of a large number of users who are present until system completion. The model would be a reasonable approximation for the servicing of a flash crowd scenario, in which a large number of users arrive almost simultaneously, which tests the scalability and efficiency of any file dissemination system most severely. However, different models may be required for other situations. Also, each user may only have a limited view of the network, and may not be able to contact users chosen uniformly at random from the entire network. We relax this last assumption using simulations in Section 6.

Finally, an important component of any peer-to-peer system is the incentive mechanism used to ensure users do not leech off the system. In this work we do not investigate incentives, but comment that it may be possible to design token-based incentive schemes that are compatible with our piece selection protocols.

# 3   Our Results: Protocols and their Performance

The main contributions of this work are outlined in this section. The piece selection protocols are described, along with corresponding performance bounds, in the following order: one-sided protocols using only pull, one-sided protocols using only push, the new hybrid one-sided protocol INTERLEAVE, and a two-sided protocol.

In our investigation of one-sided protocols, we assume that the file is divided into $k = k(n)$ pieces, where $k(n)$ is *at most* a polynomial function of $n$, and present results that hold with high probability for large $n$. Thus the relative number of pieces and users is allowed to vary over a broad range.

One-sided pull-based protocols are those where all communication occurs only via pulls, and piece selection is one-sided. Two examples within this class of protocols are:

> RANDOM PULL: In each slot each user requests a random piece from the set of pieces it does not possess.

> SEQUENTIAL PULL: Pieces are numbered $1, \ldots, k$, and in each slot each user pulls the lowest numbered piece it does not possess.

The following three theorems hold for any one-sided pull-based protocol, and for hard or soft constraints. The first is a negative result showing that the time needed to disseminate a fixed fraction of the pieces to a small fraction of users grows as the product of the number of pieces $k$ and $\log n$. This means that using one-sided pull during initial dissemination fails to exploit the potential speedup due to file splitting.

**Theorem 1** For any $0 < \beta < 1$, starting with $k$ pieces in one user each, let $T^\beta$ be the time taken till at least $\beta k$ pieces are present in $\frac{n}{\log n}$ users each, using any one-sided pull-based protocol. Then, for any $\epsilon > 0$,
$$P\left[ T^\beta \geq \beta(1-\epsilon)\, k \log n \right] \geq 1 - n^{-c}$$
for all $c > 0$, $k$ at most polynomial in $n$, and $n$ large enough.

The next theorem shows that, starting from a state where each piece is present in a fraction of the nodes, any pull-based protocol delivers all pieces to all users in $O(k + \log n)$ time, with high probability. Thus, pull finishes dissemination within a constant factor of the time needed by the optimal centralized protocol.

**Theorem 2** Let $0 < \eta < 1$ and consider any pull-based protocol. From a state such that each piece is in $\eta n$ users, if $T_\eta$ is the time till all users have all pieces then
$$P\left[ T_\eta \leq \left(\frac{\log(1+\frac{e}{\eta})}{\log(1+\frac{\eta}{e})}\right) k + \left(\frac{1+c}{\log(1+\frac{\eta}{e})}\right) \log n \right] \geq 1 - n^{-c} \qquad (1)$$

for all $c > 0$, and any $n$ and $k$.

Intuitively, the reason that pull protocols are efficient from a starting state such as the one in Theorem 2 is that each pull request has a probability greater than $\eta$ of targeting a user who can service the request.

The next theorem gives an upper bound on the completion time for a pull protocol.

**Theorem 3** Consider a network with $n$ users and $k$ pieces, each initially present at one user, implementing a pull-based protocol for piece dissemination. If $T$ is the first time that all users have all pieces, then given any $\delta > 0$, any $c > 0$, $n$ large enough, and $k$ arbitrary,

$$P\left[T \leq 4e(1+\delta)(k \log k + (1+c)k \log n)\right] \geq 1 - 2n^{-c}$$

If $k$ grows at most polynomially in $n$, then $k \log k$ is $O(k \log n)$, and Theorem 3 implies an upper bound of $O(k \log n)$. Thus, Theorems 1 and 3 together show that the completion time for any one-sided pull-based protocol is $\Theta(k \log n)$ with high probability.

One-sided push-based protocols are those where all communication occurs only via pushes, and piece selection is one-sided. An example of such a protocol is the following.

> RANDOM PUSH: In each slot each user pushes a random piece from the set of pieces it possesses.

Unlike pull-based protocols, some push-based protocols may never reach completion. For example, if pieces are pushed in a strict predefined priority order by all users, then the spread of lower priority pieces may be suppressed by the spread of the higher priority ones. However, other push-protocols such as RANDOM PUSH eventually reach completion. The following theorem shows that one-sided push-based protocols are slow in the final stages of dissemination.

**Theorem 4** For $0 < \beta < 1$, from an initial state in which $\beta k$ pieces are each absent in $\frac{n}{\log n}$ users, let $T^\beta$ be the time taken till all pieces are present in all users, using any push-based protocol. Then, for any $\epsilon > 0$,

$$P\left[T^\beta \geq \beta(1-\epsilon) k \log n\right] \geq 1 - n^{-c}$$

for all $c > 0$, $k$ at most polynomial in $n$, and $n$ large enough.

Since the completion time grows linearly in the product of $\beta k$ and $\log n$, Theorem 4 shows that purely push-based protocols provide no speedup from file splitting.

We now outline how the above results motivate the design of the hybrid efficient one-sided protocol INTERLEAVE. Theorems 1 and 4 show that any protocol relying on only one of the push or pull mechanisms can provide no speedup from file splitting. Also, pull protocols are inefficient near the start but efficient in the end, while push protocols are inefficient in the end. This indicates that it may be possible that a hybrid protocol, in which users execute pushes and pulls, can ensure efficient completion.

Theorem 2 shows that, from the viewpoint of achieving $O(k+\log n)$ dissemination time, an intermediate state – in which each piece is present in $\eta n$ users for some $\eta > 0$ independent of $n$ – is of fundamental importance. From such a state, user pulls in any hybrid protocol would enable completion in $O(k + \log n)$ time. To design a hybrid protocol whose overall completion time, from start to finish, is also $O(k + \log n)$, we would thus need to

(a) design a push protocol that reaches the intermediate state in $O(k + \log n)$ time, and

(b) combine the above push protocol with a pull-based protocol in a decentralized way.

This is the idea underlying the design of the efficient protocol INTERLEAVE.

Working towards the first of the above two objectives, we notice that while Theorem 2 guarantees the efficiency of pulls in the end, there is no analogous theorem for the efficiency of push protocols in the beginning. In particular, some push protocols may not reach the intermediate state in $O(k + \log n)$ time. We thus need to design a push protocol for this objective. Consider the one-sided push-based protocol PRIORITY PUSH.

---

PRIORITY PUSH:

- pieces are numbered $1, 2, \ldots$

- in each slot every user other than the source transmits a copy of the highest numbered piece it has received until that time.

- The source transmits piece number $i$ in the time slots $(i-1)l + 1, \ldots, il$, such that $l \geq 1$ is an integer called the *spacing* of the protocol.

---

**Theorem 5** Given any $\delta > 0$ and $0 < c < 1$, if the PRIORITY PUSH protocol with spacing $l$ is implemented, the probability that a given piece $p$ reaches $n(1 - e^{-l} - \delta)$ users within time $(1 + \delta) \log_2 n$ after leaving the source is at least $1 - 3n^{-c}$ for large enough $n$.

Before we proceed with the design of an efficient hybrid one-sided protocol, we briefly comment on the use of PRIORITY PUSH for the case when the source has the ability to generate pieces that are forward-erasure-coded versions of the original file pieces. With forward erasure coding, each user now only needs to build a large enough set of distinct coded pieces to be able to recover the original file. A protocol based on a combination of rateless forward error correction at the source, as proposed for example in [17], and piece relay within the network using PRIORITY PUSH, would work as follows: the source pushes a new coded piece in every time slot, which is labeled with a piece number as required by the PRIORITY PUSH protocol. A user at any time would transmit the highest numbered (coded) piece it currently possesses. PRIORITY PUSH ensures that each user receives approximately 63.2% of the coded pieces emerging from the source, and each of these pieces is received approximately $\log_2 n$ time slots after it emerges from the source. This means that each user can to build up a large enough collection of coded pieces in a timely fashion, enabling the decoding of the source file. Such a combination of source coding and PRIORITY PUSH may be a good candidate in scenarios such that two-way communication between users is impossible or infeasible, because in this case the pulling of pieces by users would not be possible. The delay guarantee provided by PRIORITY PUSH means that it might also be a good candidate for the relaying of source-encoded data that is of a streaming/real-time nature.

Turning to the design of an efficient hybrid one-sided protocol, observe that PRIORITY PUSH (with $l=1$) manages to deliver almost every piece to about $(1 - e^{-1})n$ nodes. This makes it a suitable candidate for combining with a pull protocol to finally generate a hybrid protocol. The fact PRIORITY PUSH tends to deliver pieces with lower numbers before pieces with higher numbers, suggests that a good pull protocol to combine with PRIORITY PUSH is the SEQUENTIAL PULL protocol. The protocols are combined by having users alternate between pushing and pulling. The performance guarantee of PRIORITY PUSH is more fragile than that of SEQUENTIAL PULL, and for this reason the hybrid protocol INTERLEAVE is designed so that the pulling does not interfere with the pushing. The protocol is described below.

---

INTERLEAVE:

- Pieces are numbered $1, 2, \ldots$

- In every odd time slot, the source pushes the piece with number one higher than the one it transmitted in the previous odd time slot. Every other user pushes the highest numbered piece it received in the previous odd time slots. The user may have a higher numbered piece obtained in an even time slot, but this is not the one chosen for transmission.

- In every even time slot, every user sends a pull request for the lowest numbered piece it does not already have. In this slot users do not distinguish pieces based on whether they were received in even or odd time slots.

---

**Theorem 6** If $T^{k_1}$ is the time INTERLEAVE takes to disseminate the $k_1$ lowest numbered pieces, then given any $s < \frac{1}{2}$ we have that

$$P\left[T^{k_1} \leq 9k_1 + 2(1+\epsilon)\log_2 n\right] \geq 1 - 5n^{-s}$$

for any $\epsilon > 0$, $k_1$ at most polynomial in $n$, and $n$ large enough.

The above theorem implies that, with high probability, INTERLEAVE achieves complete dissemination in time that is within a factor of nine of what an optimal fully centralized protocol could achieve. This means that it will be able to provide a significant file-splitting speedup for large networks.

The fact that users communicate pieces in rough order, and the delay guarantee of Theorem 6 for the lowest numbered pieces, suggests that users receive lower numbered pieces before higher numbered ones. This indicates that INTERLEAVE, or protocols of a similar design, would be useful in peer-to-peer networks in which the data to be disseminated is of a real-time nature.

It is interesting to contrast the above performance guarantee of INTERLEAVE with the single-piece results of Karp et. al. [9]. In that paper, the authors obtain a lower bound of $\Omega(n \log \log n)$ on the number of transmissions of the single piece that need to occur for complete dissemination, for any protocol relying on random user contacts of the kind studied in our paper. However, when there are multiple pieces, protocols can save bandwidth by pipelining across pieces. INTERLEAVE manages to do this pipelining in a way that results in $O(n)$ transmissions per piece, which is the optimal order.

We now move on to consider two-sided piece selection protocols. Users carry out pushes/pulls, but have knowledge of the target's current state. We consider an initial state where $n$ distinct pieces are present in the system, one in each user. Completion from such an initial state has been previously studied, often under the alternate title of "all-to-all communication". For this state, consider the following two-sided pull protocol:

> ADVOCATE: If the user does not already possess the target's initial piece, it downloads that piece. Else it pulls a random piece from among those present in the target but absent in the user.

In this protocol each user acts as an advocate for its initial piece. If each user is restricted to downloading at most one piece in every time slot, an optimal central protocol takes at least $n - 1$ time slots to complete. The following theorem shows that the ADVOCATE protocol completes in time very close to this optimal, with high probability.

**Theorem 7** Starting with each user having one unique piece, the ADVOCATE protocol operating under soft constraints finishes in $n + O(\log n)$ time with high probability.

In the above theorem the pre-constant 1 of $n$ is the best possible. The above theorem means that for large $n$ the fraction of wasted time slots is negligible.

# 4 One-sided Protocols

In this section we give the proofs of Theorems 1-6, which deal with one-sided protocols.

## 4.1 One-sided Pull-based Protocols

Since Theorem 1 is a lower bound on completion time, there is no loss of generality in assuming soft constraints. The idea behind the proof is to use a probabilistic counting argument to provide a lower bound on the number of pull requests needed per piece. Since at most $n$ pull requests occur in any given time slot, such a lower bound on the number of requests needed yields a lower bound on the dissemination time.

**Lemma 1** Consider a system with soft constraints, and an initial system state such that a given piece $p$ is present in only one user. Let $A$ be the number of pull requests for $p$ needed till it is present in $\frac{n}{\log n}$ users. Then given any $\epsilon > 0$ and $c > 0$,

$$P[A \leq (1-\epsilon)n \log n] < \frac{n^{-c}}{k}$$

for any $k$ that grows at most polynomially in $n$, and $n$ large enough.

**Proof of Lemma 1:** The probability of success of any pull request increases in the number of pull requests occurring strictly before it. It is thus sufficient to assume that the pull requests for piece $p$ occur in strict sequence, with no two being simultaneous. For such a sequence, $Geo(\frac{i}{n})$ pull requests for $p$ are needed before its occupancy goes from $i$ to $i+1$ users[2]. Thus $A$ is stochastically greater than or equal to the sum $B = Geo(\frac{1}{n}) + \ldots + Geo(\frac{1}{\log n})$. In turn, the probability $P[B \leq (1-\epsilon)n \log n]$ is shown in Lemma 10 in the appendix to be as small as is required by this lemma. ∎

**Proof of Theorem 1:** For a given pull protocol, let $A_p$ be the number of pull requests for a particular piece $p$ until it is present in $\frac{n}{\log n}$ users. Since there are at most $n$ pull requests in one time slot, it follows that for any time $t$,

$$P[T^\beta < t] \leq P\left[A_p < \frac{nt}{\beta k} \text{ for some piece } p\right] \leq \sum_p P\left[A_p < \frac{nt}{\beta k}\right]$$

From Lemma 1 we see that, if $n$ is large enough, choosing $\frac{nt}{\beta k} = (1-\epsilon)n \log n$ yields $\sum_p P\left[A_p < \frac{nt}{\beta k}\right] < n^{-c}$. This proves the theorem. ∎

We now turn our attention to Theorem 2. Here, we assume without loss of generality that the system is operating under hard constraints. Any user needs at most $k$ successful pull requests until its collection is complete. From the initial state of Theorem 2, the time to the next successful pull is always stochastically less than $Geo(\frac{\eta}{e})$[3]. Each user can thus be shown to complete in $O(k)$ time, and by a union bound the slowest of $n$ users can be shown to finish in $O(k + \log n)$ time. We now proceed to make the above argument formal.

---

[2] For any $0 < \alpha < 1$, $Geo(\alpha)$ represents a geometrically distributed random variable: $P[Geo(\alpha) > m] = (1-\alpha)^m$ for integer $m \geq 0$

[3] The target has the requested piece with probability $\eta$, and is not simultaneously targeted by any other user with probability $(1-\frac{1}{n})^{n-1} > \frac{1}{e}$

**Lemma 2** *For each user $i$, let $T_i$ be the first time slot that user $i$ has all $k$ pieces, given that each pull is successful with probability at least $\rho$. Then, for all times $t$,*

$$P\left[\max_{1\leq i\leq n} T_i > t\right] \leq n e^{-\theta t} \left(\frac{\rho e^\theta}{1-(1-\rho)e^\theta}\right)^k \tag{2}$$

*for $0 < \theta < \log\frac{1}{1-\rho}$, and any $n$ and $k$*

**Proof of Lemma 2:** Any node would need at most $k$ successful pull requests till its collection is complete. For user $i$, let $X_1, \ldots, X_k$ be the differences between successive times at which its pull requests are satisfied. Then, using the Markov inequality, for $\theta > 0$,

$$P(T_i > t) = P(X_1 + \ldots + X_k > t) \leq \frac{E[e^{\theta(X_1+\ldots+X_k)}]}{e^{\theta t}}$$

Now, at each time, the probability that the user's pull is successful is lower bounded by $\rho$. This means that any $X_j$ is stochastically upper bounded by a $Geo(\rho)$ random variable, irrespective of the other $X$'s. Thus,

$$E[e^{\theta X_j}|X_1 \ldots X_{j-1}] \leq \frac{\rho e^\theta}{1-(1-\rho)e^\theta}$$

for all $\theta$ such that $(1-\rho)e^\theta < 1$. This gives

$$\begin{aligned}
P(T_i > t) &\leq e^{-\theta t} E[e^{\theta(X_1+\ldots+X_k)}] \\
&= e^{-\theta t} E[e^{\theta(X_1+\ldots+X_{k-1})} E[e^{\theta X_k}|X_1 \ldots X_{k-1}]] \\
&\leq e^{-\theta t} \left(\frac{\rho e^\theta}{1-(1-\rho)e^\theta}\right) E[e^{\theta(X_1+\ldots+X_{k-1})}] \\
&\vdots \\
&\leq e^{-\theta t} \left(\frac{\rho e^\theta}{1-(1-\rho)e^\theta}\right)^k
\end{aligned}$$

Now, by the union bound,

$$P\left(\max_i T_i > t\right) \leq \sum_i P(T_i > t) \leq n e^{-\theta t} \left(\frac{\rho e^\theta}{1-(1-\rho)e^\theta}\right)^k$$

The lemma is thus proved. ∎

**Proof of Theorem 2:** Any pull request is successful with probability at least $\rho = \frac{\eta}{e}$. Setting the RHS of (2) to $n^{-c}$ gives

$$t = \left(\frac{1}{\theta}\log\left(\frac{\rho e^\theta}{1-(1-\rho)e^\theta}\right)\right) k + \left(\frac{1+c}{\theta}\right) \log n \tag{3}$$

Note that the choice of $\theta$ in (3) trades off between the coefficients of $k$ and $n$. Choosing $\theta = \log(1+\rho) < \log\frac{1}{1-\rho}$ gives

$$t = \left(1 - \frac{\log\rho}{\log(1+\rho)}\right) k + \frac{1+c}{\log(1+\rho)} \log n$$

Substituting the value of $\rho = \frac{\eta}{e}$ gives (1) and proves the theorem. ∎

If users are implementing SEQUENTIAL PULL, all that is required for equation (2) to hold is that piece number $p$ be present in $\eta n$ users by time slot number $p$, instead of being so from the beginning. This is because users pull in sequence, and so do not pull piece $p$ before time $p$. Choosing $\theta$ close to 0, with $\rho = \frac{\eta}{e}$, gives the following lemma, which is used in Section 4.3.

**Lemma 3** Consider a scenario such that piece $i$ is present in $\eta n$ users by time slot $i$, for each $i \in 1, \ldots, k$, and SEQUENTIAL PULL is implemented. For $\epsilon > 0$ there is a constant $M_\epsilon$, such that if $T$ is the time till all users have all $k$ pieces,

$$P\left[T > \left(\frac{e}{\eta} + \epsilon\right)k + (1+c)M_\epsilon \log n\right] \leq n^{-c}$$

for $c > 0$ and all $n$ and $k$.

**Proof of Lemma 3:** By the reasoning above, if $t$ depends on $\theta$ as in (3), then $P[T > t] \leq n^{-c}$ for $\rho = \frac{\eta}{e}$ and any value of $\theta < \log \frac{1}{1-\rho}$. Note now that

$$\lim_{\theta \to 0} \frac{1}{\theta} \log\left(\frac{\rho e^\theta}{1 - (1-\rho)e^\theta}\right) = \frac{1}{\rho}$$

and so given $\epsilon > 0$ there exists a $\delta > 0$ such that setting $\theta = \delta$ gives $t = \left(\frac{1}{\rho} + \epsilon\right)k + (1+c)M_\epsilon \log n$ where $M_\epsilon = \frac{1}{\delta}$ is a constant that depends on $\epsilon$ and grows large as $\epsilon$ becomes small. ∎

For the proof of Theorem 3, we first prove a stochastic upper bound on the number of pull requests for a given piece before it is present in $\epsilon n$ users. We then use this to provide an upper bound on the number of requests needed for all pieces to get to $\epsilon n$ users each. Since at least $n(1 - \epsilon)$ pulls take place in every time slot until this state is reached, an upper bound on the total number of pull requests needed provides an upper bound on the time taken for the system to reach such a state. For the remaining time to full completion, we use Theorem 2.

**Lemma 4** Let $A$ be the number of pull requests for a piece $p$ until it is present in all users. Then, for any $c > 0$ and any $k$ and $n$,

$$P[A > (4e)\, n \log k + 4e(1+c)\, n \log n] < \frac{n^{-c}}{k}$$

**Proof of Lemma 4:** If a user not having $p$ requests it from a user who has $p$, and no other request has the same target, then it is counted as a success. For any time $t$ let $N_t$ be the number of users who have $p$ at time $t$, and $a_t$ be the number of requests for $p$ in that time slot. $N_{t+1}$ is equal to $N_t$ plus the number of successes in the $a_t$ requests. Note that $N_t \leq N_{t+1} \leq 2N_t$, because each of the $N_t$ users can satisfy at most one request in one time slot. For any $\theta > 0$, the function $f(x) = \frac{1}{x^\theta}$ is convex for $x > 0$ and hence for any two positive numbers $a$ and $b$ if $b \le a \le 2b$ then

$$\frac{1}{a^\theta} - \frac{1}{b^\theta} \le -\frac{\theta}{(2b)^{\theta+1}}(a-b)$$

So, we have that

$$E\left[\frac{1}{N_{t+1}^\theta}\bigg| N_t = j, a_t\right] - \frac{1}{j^\theta} \le -\frac{\theta}{(2j)^{\theta+1}} E[N_{t+1} - j | N_t = j, a_t]$$

If $N_t = j$, the probability that any one of the $a_t$ requests for $p$ is successful is $j(1-\frac{1}{n})^{n-1} > \frac{j}{ne}$. Thus the expected number of new users satisfies $E[N_{t+1} - j | N_t = j, a_t] > \frac{a_t j}{ne}$ and hence

$$E\left[\frac{1}{N_{t+1}^\theta}\bigg| N_t = j, a_t\right] \le \frac{1}{j^\theta} - \frac{\theta}{(2j)^{\theta+1}} \frac{a_t j}{ne} = \left(1 - \frac{\theta}{2^{\theta+1} e} \frac{a_t}{n}\right)\frac{1}{j^\theta}$$
$$\le \frac{\beta^{-a_t}}{j^\theta} \qquad (4)$$

where $\beta = \exp(\frac{\theta}{en 2^{\theta+1}})$. Let $\mathfrak{F}_t$ denote the entire history till (and including) time $t$: it contains all the numbers $a_1 \ldots a_t$ as well as $N_1 \ldots N_t$. Define the quantity

$$M_t = \frac{\beta^{\sum_1^{t-1} a_s}}{N_t^\theta}$$

By (4),

$$E\left[M_{t+1}|\mathfrak{F}_t\right] = \beta^{\sum_1^t a_s} E\left[\frac{1}{N_{t+1}^\theta}\bigg|\mathfrak{F}_t\right] \le \frac{\beta^{\sum_1^t a_s}\beta^{-a_t}}{N_t^\theta} = M_t$$

so that $(M_t)$ is a nonnegative supermartingale with respect to $\mathfrak{F}_t$. Let $T_\alpha$ be the number of time slots required for $\alpha n$ successes. Then the optional sampling theorem and the fact that $M_1 = N_1 = 1$ imply that

$$1 \ge E\left[\frac{\beta^{\sum_1^{T_\alpha} a_s}}{N_{T_\alpha}^\theta}\right] \ge E\left[\frac{\beta^{\sum_1^{T_\alpha} a_s}}{n^\theta}\right]$$

For any number of requests $F$ we now have the following series of inequalities:

$$P\left[\sum_1^{T_\alpha} a_t > F\right] = P\left[\beta^{\sum_1^{T_\alpha} a_t} > \beta^F\right] \le \frac{E[\beta^{\sum_1^{T_\alpha} a_t}]}{\beta^F} \le \frac{n^\theta}{\beta^F} \qquad (5)$$

Setting the RHS of (5) to $\frac{n^{-c}}{k}$ and substituting the value of $\beta$ we get

$$F = \frac{2^{\theta+1}e}{\theta}n\log k + \frac{2^{\theta+1}e(\theta+c)}{\theta}n\log n$$

The choice of $\theta$ enables us to trade off between the constants of $n\log k$ and $n\log n$. Setting $\theta = 1$ proves the statement of the Lemma. ∎

The above lemma gives an upper bound on the number of pull requests required for any given piece before it achieves full occupancy. This can be used to provide an upper bound for the amount of time it takes until each piece has occupancy $\epsilon n$. This is done in the following lemma.

**Lemma 5** *Consider a system with $n$ users, starting with $k$ pieces present in at least one user each, implementing any pull-based protocol. Given $\epsilon > 0$, let $T_\epsilon$ be the first time that each piece is in at least $\epsilon n$ users. Then, for any $c > 0$ and any $n$ and $k$,*

$$P\left[T_\epsilon > \frac{4e}{1-\epsilon} k \log k + \frac{4e(1+c)}{1-\epsilon} k \log n\right] < n^{-c}$$

**Proof of Lemma 5:** Till time $T_\epsilon$ the number of users who have completed their piece collections is less than $\epsilon n$ and hence there are at least $n(1-\epsilon)$ pull requests in each slot. Let $A_\epsilon$ be the total number of attempts until each piece is in $\epsilon n$ users. Then, for a given time $t$, the event $T_\epsilon > t$ implies the event $A_\epsilon > n(1-\epsilon)t$.

Let $A$ be the number of pull attempts till all users have all pieces, and $A^p$ the number of pull requests for piece $p$ till it is present in all users. We now have that

$$P[T_\epsilon > t] \leq P[A_\epsilon > n(1-\epsilon)t] \leq P[A > n(1-\epsilon)t] \leq \sum_p P\left[A^p > \frac{n(1-\epsilon)t}{k}\right]$$

Where the last inequality is a union bound over all packets. If we choose

$$t = \frac{4e}{1-\epsilon} k \log k + \frac{4e(1+c)}{1-\epsilon} k \log n$$

then Lemma 4 implies that $P\left[A^p > \frac{n(1-\epsilon)t}{k}\right] \leq \frac{n^{-c}}{k}$ and hence $P[T_\epsilon > t] \leq n^{-c}$, completing the proof. ∎

We have already seen that pull-based protocols take at most $O(k + \log n)$ time to get from a state where each piece is in $\epsilon n$ users to one in which all users have all pieces. This, in conjunction with the Lemma 5, enables us to prove Theorem 3.

**Proof of Theorem 3:** Let $\epsilon = \frac{\delta}{2+\delta}$, and $T_\epsilon$ be the first time that each piece is present in at least $\epsilon n$ users. By Lemma 5,

$$P\left[T_\epsilon > 4e\left(1 + \frac{\delta}{2}\right)(k \log k + (1+c)k \log n)\right] \leq n^{-c}$$

Also, for large enough $n$, Theorem 2 yields that

$$P\left[T - T_\epsilon > 2e\delta (k \log k + (1+c)k \log n)\right] \leq n^{-c}$$

Putting the two equations above together proves the theorem. ∎

## 4.2 One-sided Push-Based Protocols

**Proof of Theorem 4:** The proof of Theorem 4 about the inefficiency of any one-sided push protocol in the final stages is similar to the proof of Theorem 1. Lemma 6 below is analogous to Lemma 1 for pull, and is proved in a similar fashion. It leads to the proof of Theorem 4 in the same way that Lemma 1 leads to the proof of Theorem 1. ∎

**Lemma 6** Consider an initial system state such that a given piece $p$ is absent in $\frac{n}{\log n}$ users. Let $A$ be the number of pushes for $p$ needed till it is present in all users. Then given any $\epsilon > 0$ and $c > 0$, for any pull-based protocol,

$$P[A \leq (1-\epsilon)n \log n ] < \frac{n^{-c}}{k}$$

for $k$ at most polynomial in $n$ and $n$ large enough.

**Proof of Lemma 6:** Since we are interested in the number of push transmissions for a given piece $p$, we can assume that they happen in strict sequence. Some of the transmissions occur to users already possessing $p$, and thus are not successful. For $1 \leq i \leq \frac{n}{\log n}$, let $X_i$ be the number of transmissions occurring when exactly $i$ users do not have the piece. It is easy to see that $X_i \sim GEOM(\frac{i}{n})$, and we are interested in obtaining a lower bound for $A = \sum_i X_i$. However, this is *exactly* what is done in Lemma 1, and we refer the reader to the proof of that lemma. ∎

### The PRIORITY PUSH Protocol

In the remainder of this section we describe the classical random gossip process defined in [10], give a new concentration result for this process, and use the result to prove Theorem 5, regarding the PRIORITY PUSH protocol. The random gossip process concerns $n$ users and only one message, which is initially present with one user. Users with messages are called *informed*. In every time slot, every informed user contacts another user chosen uniformly at random from the set of all users and sends (pushes) the message to this user, who is then also informed. Let $Y_t$ be the number of informed users at time $t$, and call the process $(Y_t : t \geq 0)$ the *classical gossip process*. The initial condition is $Y_0 = 1$.

We present a new concentration result for the process $Y$. Related results are given in [10, 11], but the result here is more exacting regarding the time that the message reaches the users. Let $G(y) = y + (n-y)(1-(1-\frac{1}{n})^y)$ for $y \in [1, n]$. For brevity, the dependence of $G$ on $n$ is suppressed. Note that for $y \in \{1, \ldots, n\}$,

$$G(y) = E[Y_{t+1} - Y_t | Y_t = y] \tag{6}$$

Define the deterministic sequence $(\overline{Y}_t : t \in \mathbb{Z})$ recursively by $\overline{Y}_t = 0$ for $t < 0$, $\overline{Y}_0 = 1$, and $\overline{Y}_{t+1} = G(\overline{Y}_t)$ for $t \geq 0$. The following proposition is proved in the appendix.

**Proposition 1 (Deterministic nature of the classical gossip)** *Let $0 < c < 1$, $l' \in \mathbb{Z}_+$, and $\epsilon > 0$. Then for $n$ sufficiently large:*
*(i) $\overline{Y}_{(1+\epsilon)\log_2 n - l'} \geq (1-\frac{\epsilon}{2})n$,*
*(ii) If, also, $\epsilon \leq \frac{1-c}{15}$, then*

$$P\{|Y_t - \overline{Y}_t| < 2^t n^{-2\epsilon} \text{ for } 0 \leq t \leq (1+\epsilon)\log_2 n\} \geq 1 - n^{-c},$$

*(iii) $P\{Y_{(1+\epsilon)\log_2 n} > (1-\epsilon)n\} \geq 1 - n^{-c}$.*

Thus, with high probability, the classical gossip process closely follows the deterministic sequence $\overline{Y}_t$, and reaches $n(1 - \epsilon)$ users in $(1 + \epsilon)\log_2 n$ time.

**Proof of Theorem 5:** Consider a particular piece $p$ during execution of the PRIORITY PUSH protocol, and let the time axis be adjusted so that the source first transmits $p$ in time slot 1. Note that the spread of piece $p$ and subsequent pieces is not affected by the spread of pieces preceding $p$. For any time $t$, let $A_t$ be the number of users transmitting $p$ in time slot $t$. We are interested in the process $A$, and for this purpose all higher numbered pieces are equivalent, because they cause similar interference to the spread of $p$. S, let $B_t$ be the number of users transmitting higher numbered pieces in time slot $t$. At any time, a user may be counted either in $A$, in $B$, or as not transmitting pieces that are numbered $p$ or higher. It is clear that:

- The process $(A_t + B_t : t \geq 0)$ is stochastically identical to $Y$ delayed by one time unit: $(A_t + B_t : t \geq 0) \stackrel{d}{=} (Y_{t-1} : t \geq 1)$[4] and in particular $A_t + B_t \stackrel{d}{=} Y_{t-1}$ for each $t \geq 1$.

- The process $B$ is stochastically identical to $Y$ delayed by $l + 1$ time units: Adopting the convention that $Y_t = 0$ for $t < 0$, we have $(B_t : t \geq 1) \stackrel{d}{=} (Y_{t-l-1} : t \geq 1)$.

Since $A_t = (A_t + B_t) - B_t$, the above shows that the process $A$ is the difference of two time shifted versions of the classical gossip process. Based on this idea, we shall apply Proposition 1 to deduce bounds on $A_t$. Each time $p$ is pushed, we call the node that $p$ is pushed to, the target. Thus, during the execution of the algorithm, a sequence of targets is generated, consisting of random variables that are independent, with each variable uniformly distributed over all the nodes. Even though $p$ may be pushed only a finite number of times, we can extend the target sequence if necessary, so that it is an infinite sequence of independent random variables, each uniformly distributed over the set of all $n$ nodes.

For some $\epsilon > 0$ (to be chosen later), let $T = (1 + \epsilon)\log_2 n$, and define the following events:

$$\mathcal{E}_1 = \left\{ |A_t + B_t - \overline{Y}_{t-1}| < 2^{t-1}n^{-2\epsilon} \text{ for } 1 \leq t \leq T - 1 \right\},$$
$$\mathcal{E}_2 = \left\{ |B_t - \overline{Y}_{t-l-1}| < 2^{t-l-1}n^{-2\epsilon} \text{ for } 1 \leq t \leq T - 1 \right\},$$

and let $\mathcal{E}_3$ denote the event that the first $nl(1 - \epsilon)$ targets of the target sequence includes at least $(1 - e^{-l} - \delta)n$ distinct nodes.

By Proposition 1.(ii), $P[\mathcal{E}_1 \cap \mathcal{E}_2] \geq 1 - 2n^{-c}$ for $n$ large enough. The probability of $\mathcal{E}_3$ is the same as the probability that $nl(1 - \epsilon)$ balls thrown independently and uniformly into $n$ bins cover at least $(1 - e^{-l} - \delta)n$ bins. A standard Poisson comparison argument shows that if $\epsilon$ is sufficiently small (depending on $\delta$), then $P[\mathcal{E}_3] \geq 1 - n^{-c}$ for $n$ large enough. Thus, for such $\epsilon$, $P[\mathcal{E}_1 \cap \mathcal{E}_2 \cap \mathcal{E}_3] \geq 1 - 3n^{-c}$ for $n$ large enough. It thus remains to show that, on the event $\mathcal{E}_1 \cap \mathcal{E}_2 \cap \mathcal{E}_3$, message $p$ reaches at least $(1 - e^{-l} - \delta)n$ nodes.

On the event $\mathcal{E}_1 \cap \mathcal{E}_2$, for $1 \leq t \leq T - 1$,

$$|A_t - (\overline{Y}_{t-1} - \overline{Y}_{t-1-l})| \leq |A_t + B_t - \overline{Y}_{t-1}| + |B_t - \overline{Y}_{t-l-1}| < 2^t n^{-2\epsilon}$$

---

[4]The symbol $\stackrel{d}{=}$ denotes equality in distribution

which means that
$$A_t > \overline{Y}_{t-1} - \overline{Y}_{t-l-1} - 2^t n^{-2\epsilon}. \tag{7}$$

Summing each side of (7) over $1 \leq t \leq T-1$, canceling like terms, and applying Proposition 1.(i) with $l' = l + 1$ yields

$$\sum_{t=1}^{T-1} A_t > (\sum_{t=T-l}^{T-1} \overline{Y}_{t-1}) - 2^T n^{-2\epsilon} \geq nl\left(1 - \frac{\epsilon}{2}\right) - n^{1-\epsilon} > nl(1-\epsilon),$$

for sufficiently large $n$. Now, $\sum_{t=1}^{T-1} A_t$ is the total number of times $p$ is pushed before time $T$. Thus, on the event $\mathcal{E}_1 \cap \mathcal{E}_2$, there are at least $nl(1-\epsilon)$ pushes before time $T$, and on the event $\mathcal{E}_3$, that is enough pushes to reach at least $(1 - e^{-1} - \delta)n$ nodes. Thus, $p$ does reach at least $(1 - e^{-l} - \delta)n$ nodes by time $T$, on the event $\mathcal{E}_1 \cap \mathcal{E}_2 \cap \mathcal{E}_3$. The proof of Theorem 5 is complete.

## 4.3 INTERLEAVE

In this section we analyze the performance of INTERLEAVE, and prove Theorem 6. The following two observations about INTERLEAVE facilitate its analysis:

- The pulling does not interfere with the pushing: if the system were sampled only in the odd time slots, the pieces pushed would be identical to an alternate system running only PRIORITY PUSH. In particular, the pushing of higher numbered pieces is unaffected by the spread of lower numbered pieces.

- Within the pulling in the even time slots, the spread of higher numbered pieces does not interfere with the pulling of lower numbered pieces.

Call a piece *failed* if it does not reach $\frac{ne}{5}$ users within time $2(1+\epsilon)\log_2 n$ of being pushed by the source. Note that $\frac{e}{5} < 1 - e^{-1}$ and hence, by Theorem 5 with $l = 1$, the PRIORITY PUSH operating in the odd time slots ensures that $P[p \text{ fails}] < n^{-c}$ for $0 < c < 1$, and $n$ large enough.

For each $i \in 1, \ldots, k$ define $T^i = \min\{t : \text{each piece in } 1, \ldots, i \text{ present in all users }\}$. We are interested in finding an upper bound on any given $T^i$. So for the remaining analysis in this section we choose some $k_1 \leq k$ and provide an upper bound on $T^{k_1}$.

**Lemma 7** Given $1 \leq q_1 < \ldots < q_m$, let $\mathcal{E}$ be the event that $\{q_1, \ldots, q_m\}$ is the set of all failed pieces numbered less than or equal to $k_1$. Then, for any $c > 0$ and $n$ large enough,

$$P\left[T^{k_1} > \begin{pmatrix} 8.8k_1 + 2(1+\epsilon)\log_2 n \\ + m\gamma(1+c)(\log n + \log m) \end{pmatrix} \middle| \mathcal{E}\right] \leq 2n^{-c}$$

such that $\gamma$ is a constant independent of $n$

**Proof of Lemma 7:** For each $i \in 1, \ldots, k$ let $T^i$ be as above and let
$$\widehat{T}^i = \max\left\{T^i, \; 2i + 2(1+\epsilon)\log_2 n\right\}$$
Consider any two successive failed pieces $q_j$ and $q_{j+1}$, and let $q_{j+1} - q_j = l$. Then, for each $1 \leq s \leq l-1$, piece number $q_j + s$, which has not failed, is present in at least $n(1 - e^{-1} - \epsilon)$ users by time $\widehat{T}^{q_j} + 2s$. Also, after $\widehat{T}^{q_j}$, all users pull pieces numbered $q_j + 1$ or higher. For this scenario, Lemma 3 implies that all users obtain all pieces $i$ such that $q_j < i < q_{j+1}$ within
$$\left(\frac{1}{\rho} + \epsilon\right)(q_{j+1} - q_j - 1) + M_\epsilon (1+c)\log n$$
pull slots, with probability at least $1 - n^{-c}$, where $\rho = \frac{1 - e^{-1} - \epsilon}{e}$. Choosing $\epsilon$ so small that $\frac{1}{\rho} + \epsilon < 4.4$, this implies that
$$\widehat{T}^{q_{j+1}-1} \leq \widehat{T}^{q_j} + 8.8(q_{j+1} - q_j) + 2M_\epsilon(1+c)\log n$$
with probability at least $1 - n^{-c}$. Also, by time $\widehat{T}^{q_{j+1}-1}$ all users pull for pieces $q_{j+1}$, or higher pieces if they already have $q_{j+1}$. By Theorem 3 with $k=1$,
$$\widehat{T}^{q_{j+1}} - \widehat{T}^{q_{j+1}-1} \leq 8e(1+\delta)(1+c)\log n$$
with probability at least $1 - n^{-c}$. The last two inequalities above imply that
$$\widehat{T}^{q_{j+1}} \leq \widehat{T}^{q_j} + 8.8(q_{j+1} - q_j) + \gamma(1+c)\log n \tag{8}$$
with probability at least $1 - 2n^{-c}$, such that $\gamma = 8e(1+\delta) + 2M_\epsilon$. Defining $\widehat{T}^{q_0} = 2(1+\epsilon)\log_2 n$ and $q_{m+1} = k_1$, and summing (8) over all $j$ shows that
$$\widehat{T}^{k_1} \leq 8.8k_1 + 2(1+\epsilon)\log_2 n + m\gamma(1+c)\log n$$
with probability at least $1 - 2mn^{-c}$. Replacing $c$ by $c + \frac{\log m}{\log n}$ proves the lemma. ∎

We are now ready to prove the performance guarantee of the INTERLEAVE protocol as given in Theorem 6.

**Proof of Theorem 6:** Let $m$ be the number of failed pieces in $\{1, \ldots, k_1\}$. Then, by the Markov inequality we have that
$$P[m > k_1 n^{-s}] \leq \frac{E[m]}{k_1 n^{-s}}$$
By Theorem 5, if $n$ is large enough the probability piece $p$ fails is less than $3n^{-2s}$ for $s < \frac{1}{2}$. This means that $E[m] \leq 3k_1 n^{-2s}$ and so
$$P[m > k_1 n^{-s}] \leq 3n^{-s} \tag{9}$$
Now, in Lemma 7, if $m \leq k_1 n^{-s}$ then the fact that $k_1$ is at most polynomial in $n$ implies that $m\gamma(1+c)(\log n + \log m)$ is $o(k_1)$, so for $n$ large enough its value is less than $(0.2)k_1$. So Lemma 7 yields:
$$P\left[T^{k_1} > 9k_1 + 2(1+\epsilon)\log_2 n \mid m \leq k_1 n^{-s}\right] \leq 2n^{-s} \tag{10}$$
Theorem 6 follows from (9) and (10). ∎

# 5 Two-sided Protocols

Consider the initial system condition where there are $n$ users in the system, each possessing a unique piece. The initial piece with user $i$ is denoted by $p_i$. In every time slot each user contacts a random other user to request a piece. The piece selection is two-sided, and is made according to the ADVOCATE protocol. To prove Theorem 7, we analyze the evolution of the system by breaking time up into phases.

At any time $t$ define a user $i$'s *primary collection* to be the set of pieces $p_j$ such that $i$ contacted user $j$ by time $t$. Note that a piece $p_j$ can be in the primary collection of user $i$ even if $i$ did not get $p_j$ directly from user $j$. Note that at any time, the primary collections of the users are independent of each other. The pieces that user $i$ has that are not in its primary collection are said to be in user $i$'s *secondary collection*.

**Phase I**

This phase goes from the beginning up until time $n^{\frac{1}{4}-\delta}$, for some fixed $0 < \delta < \frac{1}{4}$. During this phase, with high probability, no user contacts the same user twice. This guarantees that all users are successful in each slot in this period.

**Lemma 8** *For $\delta < \frac{1}{4}$, at time $n^{\frac{1}{4}-\delta}$ all users have at least $n^{\frac{1}{4}-\delta} - 1$ pieces each, with probability at least $1 - n^{-4\delta}$*

**Proof of Lemma 8:** A user $i$ is said to *repeat* in time slot $s$ if it contacts a user in slot $s$ that it had contacted previously. Given a time $t = n^{\frac{1}{4}-\delta}$, the probability of a given user repeating in any given time slot in $1, \ldots, t$ is less than $\frac{t}{n}$. Thus the probability of a given user repeating twice or more by time slot $t$ is less than $\binom{t}{2}(\frac{t}{n})^2$, which is less than $\frac{t^4}{n^2}$. Taking a union bound over the set of all users, we get that

$$P[\text{any user repeats at least twice by time } t] \leq n\frac{t^4}{n^2} = n^{-4\delta}$$

Finally, if every user repeats at most once, then every user misses at most one piece by time $n^{\frac{1}{4}-\delta}$. ∎

**Phase II**

This phase continues up until time $\frac{n}{2}$. Beyond time $n^{\frac{1}{4}-\delta}$, the users start repeating contacts more often, and hence the technique of Phase I is not applicable. For Phase II, we make use of the fact that the sizes of the user's primary collections are large enough to ensure that useful pieces can still be found in these primary collections.

For two users $A$ and $B$, let $P_A$ and $P_B$ denote their respective primary collections, and $S_A$ and $S_B$ their secondary collections, at some time $t \geq n^{\frac{1}{4}-\delta}$.

**Lemma 9** *For any $\epsilon > 0$ if $P_B - P_A$ denotes the set of pieces in $P_B$ but not in $P_A$ then*

$$P\left[|P_B - P_A| < n(1 - e^{-\frac{t}{n}})e^{-\frac{t}{n}}(1-\epsilon)\right] \leq 4\,e^{-\epsilon^2 n^{\frac{1}{4}-\delta}/9}$$

$$P\left[|P_A| < n(1 - e^{-\frac{t}{n}})(1-\epsilon)\right] \leq 2e^{-\epsilon^2 n^{\frac{1}{4}-\delta}/9}$$

**Proof of Lemma 9:** Each of the users $A$ and $B$ make $t$ contacts in time $t$. Consider now an alternate system where $A$ and $B$ make a random number of contacts $N_A$ and $N_B$ in time $t$, which are independent and distributed according to $Poisson(t)$. Denote the primary and secondary collections of the users in the alternate system with a hat. Observe that $\widehat{P}_A$ and $\widehat{P}_B$ are stochastically increasing in $N_A$ and $N_B$ respectively, and so we have that for any $x > 0$,

$$\begin{aligned}
P[|P_B - P_A| < x] &< P\left[|\widehat{P}_B - \widehat{P}_A| < x \mid N_B \leq t \text{ and } N_A \geq t\right] \\
&= \frac{P[|\widehat{P}_B - \widehat{P}_A| < x \text{ and } N_B \leq t \text{ and } N_A \geq t]}{P[N_B \leq t \text{ and } N_A \geq t]} \\
&\leq \frac{P[|\widehat{P}_B - \widehat{P}_A| < x]}{P[N_B \leq t \text{ and } N_A \geq t]} \\
&\leq 4\,P[Bin(n, e^{-\frac{t}{n}}(1 - e^{-\frac{t}{n}})) < x]
\end{aligned}$$

where for the last inequality we have used the fact that for the Poisson-distributed random variables $N_A$ and $N_B$, $P[N_A \geq t] \geq \frac{1}{2}$ and $P[N_B \leq t] \geq \frac{1}{2}$.

Setting $x = n(1 - e^{-\frac{t}{n}})e^{-\frac{t}{n}}(1-\epsilon)$, the Chernoff Bound on the binomial distribution (see for example Theorem 4.5 of [18]) gives

$$P\left[|P_B - P_A| < n(1 - e^{-\frac{t}{n}})e^{-\frac{t}{n}}(1-\epsilon)\right] \leq 4\,e^{-n(1-e^{-\frac{t}{n}})e^{-\frac{t}{n}}\epsilon^2/3}$$

Now, for $n^{\frac{1}{4}-\delta} \leq t \leq \frac{n}{2}$,

$$n(1 - e^{-\frac{t}{n}})e^{-\frac{t}{n}} \geq t(1 - e^{-1})e^{-\frac{1}{4}} > \frac{n^{\frac{1}{4}-\delta}}{3}$$

This proves the first part of the lemma. The second part is proved along similar lines:

$$\begin{aligned}
P[|P_A| < x] &\leq P\left[|\widehat{P}_A| < x \mid N_A \leq t\right] \leq 2P\left[|\widehat{P}_A| < x\right] \\
&= 2P\left[Bin(n, 1 - e^{-\frac{t}{n}}) < x\right]
\end{aligned}$$

Setting $x = n(1 - e^{-\frac{t}{n}})(1-\epsilon)$ and using the Chernoff bound as above proves the second part of the Lemma. ∎

Now, $|P_A + S_A| \leq t$, and hence the second part of Lemma 9 implies that with high probability $|S_A| < t - n(1 - e^{-\frac{t}{n}})(1 - \epsilon)$.

Now, suppose at time $t$ user $A$ contacts user $B$. If $|S_A| < |P_B - P_A|$ then $A$ is able to receive a new piece from $B$ (although this condition is not necessary). From the above arguments, it follows that if $t - n(1 - e^{-\frac{t}{n}})(1 - \epsilon) < n(1 - e^{-\frac{t}{n}})e^{-\frac{t}{n}}(1 - \epsilon)$, or equivalently if $\frac{t}{n} < (1 - \epsilon)(1 - e^{-\frac{2t}{n}})$, then $A$ can receive a piece from $B$ with failure probability less than $6\,e^{-\epsilon^2 n^{\frac{1}{4}-\delta}/9}$. For small enough $\epsilon$, it can be shown that all users are successful with probability $n^2 e^{-0.05 n^{\frac{1}{4}-\delta}}$ for all times up until time $\frac{n}{2}$.

Thus, in conjunction with the results of the first phase, it can be shown that all users have at least $\frac{n}{2} - 1$ pieces after $\frac{n}{2}$ time slots.

**Phase III**

This phase starts from time $\frac{n}{2}$ and ends at time $n$. In this phase the primary collections may not be large enough to guarantee the existence of a useful piece among every pair of users. So, for the third phase, we need to leverage the spread of pieces across users.

Let $A$ be a set of pieces of size $L = 8 \log n$, and let $\epsilon = \frac{3 \log n}{n}$. $A$ is said to be a "bad" set if the number of users having no piece in $A$ at time $\frac{n}{2}$ is greater than $n\epsilon$.

For any given user, the probability that it has no piece in $A$ is less than or equal to the probability that user does not have a piece of $A$ in its primary collection. This probability is further less than or equal to $(1 - \frac{L}{n})^{\frac{n}{2}}$, because the user has made $\frac{n}{2}$ contacts. Further, $(1 - \frac{L}{n})^{\frac{n}{2}} < e^{-\frac{L}{2}}$. Thus, the number of users who do not have any pieces in $A$ at time $\frac{n}{2}$ is stochastically smaller than a $Binomial(n, e^{-\frac{L}{2}})$ random variable. Thus,

$$P[\text{given set } A \text{ is bad}] \leq P\left[Binomial(n, e^{-\frac{L}{2}}) > n\epsilon\right]$$

Since there are $\binom{n}{L}$ possible sets of size $L$, a union bound over all such sets of size $L$ gives

$$P[\text{there exists a bad set of size } L] \leq \binom{n}{L} P\left[Binomial(n, e^{-\frac{L}{2}}) > n\epsilon\right]$$

$$\leq \binom{n}{L}\binom{n}{\epsilon n} e^{-nL\epsilon/2}$$

$$\leq \frac{n^L}{L!} \frac{n^{\epsilon n}}{(\frac{\epsilon n}{e})^{\epsilon n}} e^{-nL\epsilon/2}$$

$$\leq \exp\left(L \log n + n\epsilon + n\epsilon \log \frac{1}{\epsilon} - \frac{nL\epsilon}{2}\right)$$

The second inequality above was obtained from the relation $P[Binomial(n, p) \geq k] \leq \binom{n}{k} p^k$.

Now, substituting the value of $L$ and $\epsilon$ makes the exponent in the last inequality equal to

$$8(\log n)^2 + 3 \log n + 3(\log n)^2 - 3(\log n)(\log(3 \log n)) - 12(\log n)^2$$

which is less than $-(\log n)^2$ for $n \geq 3$. Thus, we have

$$P[\text{there exists a bad set of size } 8\log n] < e^{-(\log n)^2}$$

This means that any user missing at least $L = 8\log n$ pieces fails to obtain a useful piece with probability at most $\epsilon = \frac{3\log n}{n}$. So, at time $n$, the probability that a given user has $10\log n$ pieces missing is less than $P[Binomial(\frac{n}{2}, \frac{3\log n}{n}) > 10\log n]$. Using a Chernoff bound on the binomial distribution (in particular, part 3 of Theorem 4.4 in [18]) it can be shown that the probability that a given user has $10\log n$ pieces missing is less than $2^{-10\log n}$, which is less than $n^{-6}$. Taking a union bound over the set of users, we can show that all users have at least $n - 10\log n$ pieces by time $n$ with high probability.

**Phase IV**

This phase begins at time $n$ and finishes when every user has received every piece. In this phase each of the users has at most $10\log n$ pieces left to finish. Each of these pieces is present in at least $n(1 - e^{-1} - \epsilon)$ of the network. We can now apply Theorem 2 with $k = 10\log n$ and $\eta = 1 - e^{-1} - \epsilon$ to conclude that all users finish in $n + O(\log n)$ time with high probability.

# 6 Simulations

In this section we investigate the performance of the PRIORITY PUSH and INTERLEAVE protocols using simulations. In the system model analyzed in this paper each user has the ability to communicate with another user chosen at random from the entire network. A more realistic assumption might be to let each user have only a limited view of the network that does not change over time. Specifically, we assume that each user has a fixed list of a small number of other users, which we shall refer to as its *contact list*. A user only pushes to and pulls from other users in its contact list. Each contact list is generated randomly, independent of other contact lists. It remains constant for all time.

Consider now the case that users implement INTERLEAVE, but in each time slot a user communicates with a neighbor chosen at random from within its contact list. The source however still pushes pieces in every other time slot to another user chosen uniformly at random from the set of all users. Figure 1 displays the observed time taken for $k = 1000$ pieces to be disseminated to $n = 500$ users, versus the size of the contact lists. From this we can see that if each user has a contact list of size 8 or more, the completion time using INTERLEAVE is close to $2(k + \log_2 n) \approx 2020$, which is much better than the $10k + 2\log_2 n$ predicted by Theorem 6. This difference suggests that the proof technique for Theorem 6 is conservative.

Besides the overall completion time, we are also interested in the time a typical piece takes to get from the source to a typical user. Specifically, if a piece $i$ emerges from the source at time $t$ and reaches a user $j$ at time $t + d$, we say that delay$(i, j) = d$. The *average delay profile* $D(d)$ is the average of $\mathbf{1}_{\text{delay}(i,j) \leq d}$ over all possible choices of user $i$ and piece $j$:

$$D(d) \;=\; \frac{1}{nk} \sum_{i=1}^{n} \sum_{j=1}^{k} \mathbf{1}_{\text{delay}(i,j) \leq d}$$

where $\mathbf{1}_{\text{delay}(i,j) \leq d}$ is 1 if and only if $\text{delay}(i,j) \leq d$ and 0 otherwise.

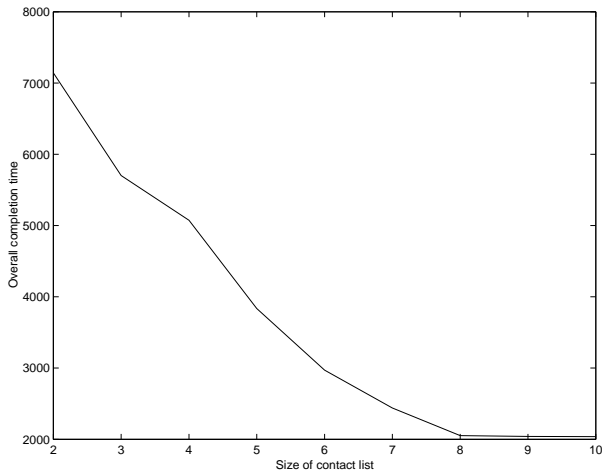

Figure 1: The completion times of INTERLEAVE for $n = 500$ users and $k = 1000$ pieces, versus the size of the user contact lists.

Different piece selection protocols have different average delay profiles. Also, for a given dissemination protocol, the average delay profile can vary with the size of each user's contact list. A delay profile rising further to the left implies, on average, faster dissemination of pieces. Figure 2 plots the average delay profiles for $k = 1000$ pieces being disseminated to $n = 500$ users, for different choices $m$ of user contact list size. From the figure we see that users having contact lists of size two leads to poor performance, but with four or five contacts the average delay profile is comparable to that achieved if each user were to have a complete view of the network.

We now turn our attention to the PRIORITY PUSH protocol. Figure 3 plots the average delay profiles for different choices of the spacing $l$, if each user has an entire view of the network. Recall that the source transmits a new piece every $l$ time slots. The final limiting value of each delay profile represents the final fraction of users a typical piece reaches. This is equivalent to the fraction of pieces a typical user ultimately receives. As predicted by Theorem 5, a spacing of $l$ has a limiting value of approximately $1 - e^{-l}$.

# 7  Discussion

In this work we

- investigated the speedup achieved in file dissemination by breaking the file into pieces.

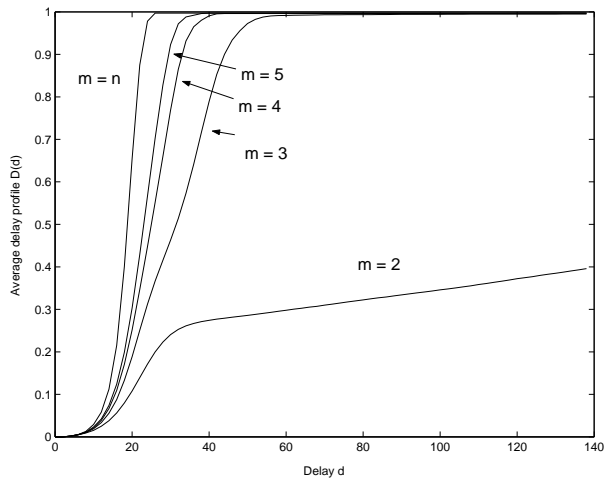

Figure 2: The average delay profiles of INTERLEAVE for $n = 500$ users and $k = 1000$ pieces, for different sizes $m$ of user contact lists.

- investigated the performance of one-sided piece selection protocols,
- designed the efficient piece selection protocol INTERLEAVE,
- illustrated why the PRIORITY PUSH protocol would be useful for the relay of source-coded pieces, and
- designed an efficient two-sided protocol for all-to-all exchange.

We believe that the techniques and results of this work will aid in the understanding of systems that involve the decentralized dissemination of large files. We would like to emphasize that the dissemination of multiple pieces of data over unstructured networks is significantly different from the dissemination of a single piece, but also note that insights gained from single-piece dissemination can be effectively leveraged to design protocols for multiple pieces. It would be interesting to investigate the spread of multiple pieces in unstructured networks using more detailed system models.

# 8   Appendix

## 8.1   Lemma for Purely Pull-Based Protocols

Suppose $n$ and $k$ are positive integers. Let $B_k = \sum_{i=1}^{k} X_i$, where the $X_i$ are independent, and $X_i$ has the $Geo(\frac{i}{n})$ distribution. Then $B_k$ represents the number of pulls needed (in the best case of sequential pulling) in order for $k$ nodes to acquire a packet initially in one node.

**Lemma 10** *Let $S_k$ denote the event $S_k = \{B_k \leq (1 - \epsilon)n \log n\}$. (Suppose that $(1 - \epsilon)n \log n$ is integer valued.) Then $P[S_k] \leq 2 \exp(-n^{-(1-\epsilon)} k)$.*

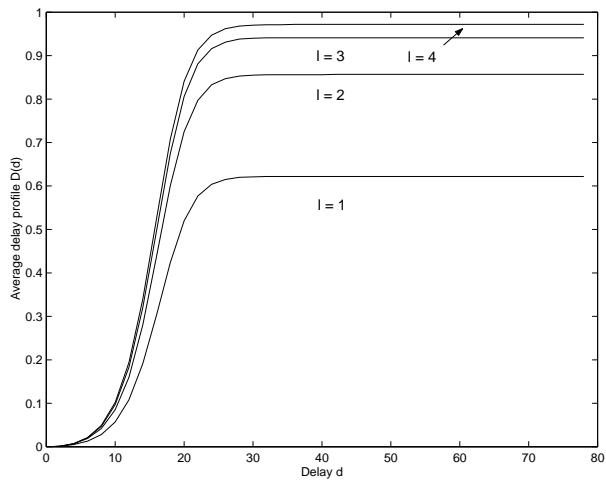

Figure 3: The average delay profiles of PRIORITY PUSH for $n = 500$ users and $k = 1000$ pieces, for different values of spacing $l$.

For example, if $k = \frac{n}{\log n}$, then $P[S_k] \leq 2\exp(-n^\epsilon/\log n)$, or, if $k = n^{1-\epsilon/2}$, then $P[S_k] \leq 2\exp(-n^{\epsilon/2})$.

**Proof of Lemma 10:** $B_k$ has the same distribution as the completion time for the coupon collection problem, starting from an initial collection of coupons that is missing $k$ types of coupons. Thus $S_k$ is the event that in a sample of $(1-\epsilon)n \log n$ random coupons, there is at least one coupon of each of the $k$ missing types. For the sake of comparison, consider the case that a random number $N$ of random coupons is examined, where $N$ is a Poisson random variable with mean $(1-\epsilon)n \log n$. Then $P\{N \geq (1-\epsilon)n \log n\} \geq 1/2$. (see [18], Exercise 5.13). So
$$P[\text{success using } (1-\epsilon)n \log n \text{ coupons}] \leq 2P[\text{success using } N \text{ coupons}].$$
For the random sample size, the numbers of coupons of different types are independent Poisson random variables with mean $(1-\epsilon)\log n$. Thus, a given type is found in the random size sample of random coupons with probability $1 - e^{(-(1-\epsilon)\log n} = 1 - n^{-(1-\epsilon)}$. Hence,
$$P[\text{success using } N \text{ coupons}] = (1 - n^{-(1-\epsilon)})^k \leq \exp(-kn^{-(1-\epsilon)})$$
The lemma is thus proved. ∎

## 8.2 Proof of Determinism of Classical Gossip Process

**Lemma 11** *The function $G$ as defined in (6) is a strictly increasing map of $[1, n]$ onto $[2-\frac{1}{n}, n]$, and it is Lipschitz continuous with Lipschitz constant 2. Furthermore,*
$$G(y) \geq 2y(1 - \frac{y}{n}) \tag{11}$$
*and*
$$n - G(y) \leq (n-y)e^{-y/n} \tag{12}$$

**Proof of Lemma 11:** Note that $G(1) = 2 - \frac{1}{n}$ and $G(n) = n$. Differentiating yields

$$G'(y) = (1 - \frac{1}{n})^y (1 - (n-y)\ln(1 - \frac{1}{n}))$$

Now $0 \leq -\ln(1-u) \leq \frac{u}{1-u}$ for $0 \leq u \leq 1$. Hence for $y \in [1, n]$,

$$0 \leq G'(y) \leq 1 - (n-1)\log(1 - \frac{1}{n}) \leq 2$$

so that the first sentence of the lemma is proved. Inequality (11) follows from the fact $(1-\frac{1}{n})^y \geq 1 - \frac{y}{n}$. The function $G$ can be expressed as

$$G(y) = y + (n-y)(1 - \underline{e}^{-y/n})$$

where $\underline{e} = (1-\frac{1}{n})^{-n}$, and (12) follows from the fact $\underline{e} \geq e$. ∎

**Lemma 12** *For $\epsilon > 0$, if $n$ is sufficiently large, $\overline{Y}_{(1+\epsilon)\log_2 n} \geq (1 - \frac{\epsilon}{2})n$.*

**Proof of Lemma 12:** If the lemma is true for some $\epsilon > 0$, then it is trivially true for larger values of $\epsilon$, so without loss of generality it can be assumed that $0 < \epsilon < 0.1$. By (11), if $1 \leq \overline{Y}_t \leq \frac{\epsilon n}{3}$, then $\overline{Y}_{t+1} \geq (2(1-\frac{\epsilon}{3}))\overline{Y}_t$. Hence,

$$\overline{Y}_t \geq \min\left\{\left(2\left(1 - \frac{\epsilon}{3}\right)\right)^t, \frac{n\epsilon}{3}\right\} \quad \text{for } t \geq 0.$$

Let $t_1 = (1 + (0.9)\epsilon)\log_2 n$. The fact $\ln(1 - \frac{\epsilon}{3}) \geq -\frac{\epsilon/3}{1-\epsilon/3} \geq -(0.35)\epsilon$ yields

$$\left(2\left(1 - \frac{\epsilon}{3}\right)\right)^{t_1} \geq \exp([\ln 2 - (0.35)\epsilon]t_1)$$
$$= n\exp\left(\{[\ln 2 - (0.35)\epsilon](0.9) - (0.35)\}\epsilon \log_2 n\right)$$
$$\geq n\exp((0.20)\epsilon \log_2 n) \geq n$$

Thus, $\overline{Y}_{t_1} \geq \min\{n, \frac{\epsilon n}{3}\} = \frac{\epsilon n}{3}$.

Similarly, if $\overline{Y}_t \leq \frac{n}{3}$, then $\overline{Y}_{t+1} \geq \frac{4}{3}\overline{Y}_t$. Hence, if $t_2 = t_1 + \ln(\frac{1}{\epsilon})/\ln(\frac{4}{3})$, then $\overline{Y}_{t_2} \geq \min\{\frac{\epsilon n}{3}(\frac{4}{3})^{t_2-t_1}, \frac{n}{3}\} = \frac{n}{3}$.

Finally, (12) yields that if $\frac{n}{3} \leq \overline{Y}_t \leq n$, then $n - \overline{Y}_{t+1} \leq (n - \overline{Y}_t)e^{-1/3}$. Hence, if $t_3 = t_2 + 3(\ln(\frac{2}{\epsilon}) - \ln(\frac{3}{2}))$, then $n - \overline{Y}_{t_3} \leq (\frac{2n}{3})e^{-\frac{1}{3}(t_3-t_2)} = \frac{n\epsilon}{2}$. That is, $\overline{Y}_{t_3} \geq n(1 - \frac{\epsilon}{2})$. If $n$ is sufficiently large, $t_3 - t_1 \leq (0.1)\epsilon \log_2 n$, so that $t_3 \leq (1+\epsilon)\log_2 n$, and the lemma follows. ∎

**Lemma 13** *Given $0 < c < 1$ and $0 < \epsilon \leq \frac{1-c}{15}$, let $t_0 = \lfloor 7\epsilon \log_2 n \rfloor$. Then for sufficiently large $n$,*

$$P\{Y_{t_o} = 2^{t_o}\} \geq 1 - \frac{1}{2}n^{-c} \tag{13}$$

$$|2^t - \overline{Y}_t| < 2^t n^{-3\epsilon} \quad \text{for} \quad 0 \leq t \leq t_o \tag{14}$$

**Proof of Lemma 13:** Although the push transmissions occur in rounds, the selections of people can be considered sequentially. If the first $2^t$ selections are distinct, then $Y_t = 2^t$. Each of these selections is distinct from the ones before it with probability at least $1 - \frac{2^t}{n}$, so

$$P[Y_t = 2^t] \geq \left(1 - \frac{2^t}{n}\right)^{2^t} \geq 1 - \frac{(2^t)^2}{n}$$

Hence, taking $t = t_o$ and using the fact $t_o \leq 7\epsilon \log_2 n + 2$,

$$P\{Y_t = 2^t : 0 \leq t \leq t_o\} = P\{Y_{t_o} = 2^{t_o}\} \geq 1 - 4n^{14\epsilon - 1} \geq 1 - \frac{1}{2}n^{-c}$$

for $n$ sufficiently large, and (13) is proved.

If $0 \leq t \leq t_o - 1$, then $\overline{Y}_t \leq n^{7\epsilon}$. This fact and (11) imply

$$2\overline{Y}_t \geq \overline{Y}_{t+1} \geq 2\overline{Y}_t\left(1 - \frac{\overline{Y}_t}{n}\right) \geq 2\overline{Y}_t(1 - n^{7\epsilon - 1})$$

Hence, for $0 \leq t \leq t_o$,

$$\begin{aligned} 2^t \geq \overline{Y}_t &\geq 2^t(1 - n^{7\epsilon - 1})^{7\epsilon \log_2 n} \\ &\geq 2^t(1 - (n^{7\epsilon - 1})7\epsilon \log_2 n) \\ &\geq 2^t(1 - n^{-3\epsilon}) \text{ for } n \text{ large enough} \end{aligned}$$

because $10\epsilon < 1$. Thus, (14) is proved. ∎

**Lemma 14** *For $t \geq 0$, $P[|Y_{t+1} - G(Y_t)| \geq Y_t n^{-3\epsilon}|Y_t] \leq \exp(-\frac{(Y_t)n^{-6\epsilon}}{2})$*

**Proof of Lemma 14:** The idea is to apply the Azuma-Hoeffding inequality [18]. In a round beginning with $Y_t$ informed people, there are $Y_t$ selections. Each selection has the potential to increase the number of informed people by one. Thus, given $Y_t$, the variable $Y_{t+1} - G(Y_{t_1})$ can be viewed as the ending value of a martingale with $Y_t$ steps, where the interval of uncertainty for each step has length one. ∎

**Proof of Proposition 1:** If Proposition 1 is true for $l' = 0$, then it is true for any $l' \in \mathbb{Z}_+$, because the term $l'$ can be covered by using a smaller value of $\epsilon$. Thus, in the proof, we can take $l' = 0$. Then Proposition 1.(i) is the same as Lemma 12, proved above. Note that if Proposition 1.(iii) holds for some $\epsilon > 0$, then it also holds for larger $\epsilon$. Thus, Proposition 1.(iii) can be proved with the additional assumption that $\epsilon \leq \frac{1-c}{15}$. Then, Proposition 1.(iii) follows easily from Proposition 1.(i) and Proposition 1.(ii). It remains to prove Proposition 1.(ii).

Let $0 < c < 1$ and $0 < \epsilon \leq \frac{1-c}{15}$. As in Lemma 13, let $t_0 = \lfloor 7\epsilon \log_2 n \rfloor$. Let $E_0$ denote the event $E_0 = \{Y_{t_o} = 2^{t_o}\}$. Lemma 13 implies that $P[E_o] \geq 1 - \frac{1}{2}n^{-c}$, and

$$E_0 \subset \{|Y_t - \overline{Y}_t| \leq 2^t n^{-3\epsilon} \text{ for } 0 \leq t \leq t_o\} \tag{15}$$

For $t_o \leq t \leq (1+\epsilon)\log_2 n$, let $E_t = \{|Y_{t+1} - G(Y_t)| \leq 2^t n^{-3\epsilon}\}$. If $E_0$ is true and $t \geq t_o$, then $Y_t \geq Y_{t_o} = 2^{t_o} \geq n^{7\epsilon}$, so by Lemma 14, $P[E_t^c|E_0] \leq \exp(-\frac{n^\epsilon}{2})$ for $t \geq t_o$. Therefore, with $E = E_0 \cap \left( \bigcap_{t_o \leq t \leq (1+\epsilon)\log_2 n} E_t \right)$,

$$P[E^c|E_0] \leq \sum_{t=t_o}^{(1+\epsilon)\log_2 n} P[E_t^c|E_0] \leq \sum_{t=t_o}^{(1+\epsilon)\log_2 n} \exp(-\frac{n^{-\epsilon}}{2})$$

$$\leq (\log_2 n) \exp(-\frac{n^{-\epsilon}}{2}) \leq \frac{1}{2} n^{-c}$$

for $n$ large enough. Thus, $P[E] = (1 - P[E^c|E_0])P[E_o] \geq (1 - \frac{1}{2}n^{-c})^2 \geq 1 - n^{-c}$.

Let $F_t = \{|Y_t - \overline{Y}_t| < 2^t n^{-2\epsilon}\}$. It remains to show that $E \subset F_t$ for $1 \leq t \leq (1+\epsilon)\log_2 n$. Lemma 13.(ii) implies that $E \subset E_0 \subset F_t$ for $0 \leq t \leq t_o$. So let $t_o + 1 \leq t \leq (1+\epsilon)\log_2 n$. Let $G^k$ denote the composition of the function $G$ with itself $k$ times. Expressing $Y_t$ as a telescoping sum yields

$$Y_t = G^{t-t_o}(Y_{t_o}) + \sum_{i=t_o}^{t-1} G^{t-i-1}(Y_{i+1}) - G^{t-i-1}(G(Y_i))$$

and the definition of $\overline{Y}_t$ yields $\overline{Y}_t = G^{t-t_o}(\overline{Y}_{t_o})$. Thus,

$$|Y_t - \overline{Y}_t| \leq |G^{t-t_o}(Y_{t_o}) - G^{t-t_o}(\overline{Y}_{t_o})| + \sum_{i=t_o}^{t-1} |G^{t-i-1}(Y_{i+1}) - G^{t-i-1}(G(Y_i))|$$

Now $G^k$ is Lipschitz continuous with Lipschitz constant $2^k$. On the event $E$, by (15), $|Y_{t_o} - \overline{Y}_{t_o}| \leq 2^{t_o} n^{-3\epsilon}$ and, because $E \subset E_t$, $|Y_{t+1} - G(Y_t)| \leq 2^t n^{-3\epsilon}$ for $t_o \leq t \leq (1+\epsilon)\log_2 n$. Therefore,

$$|Y_t - \overline{Y}_t| \leq 2^{t-t_o}(2^{t_o} n^{-3\epsilon} + \sum_{i=t_o}^{t-1} 2^{t-i-1} 2^i n^{-3\epsilon}$$

$$\leq 2^t n^{-3\epsilon} + (t-t_o)2^{t-1} n^{-3\epsilon} \leq 2^t n^{-2\epsilon},$$

assuming $n$ is so large that $(\log_2 n)n^{-\epsilon} \leq 1$. Therefore, if $n$ is sufficiently large, $E \subset F_t$ for $0 \leq t \leq (1+\epsilon)\log_n$. This establishes Proposition 1.(ii), and the entire proposition is proved. ∎

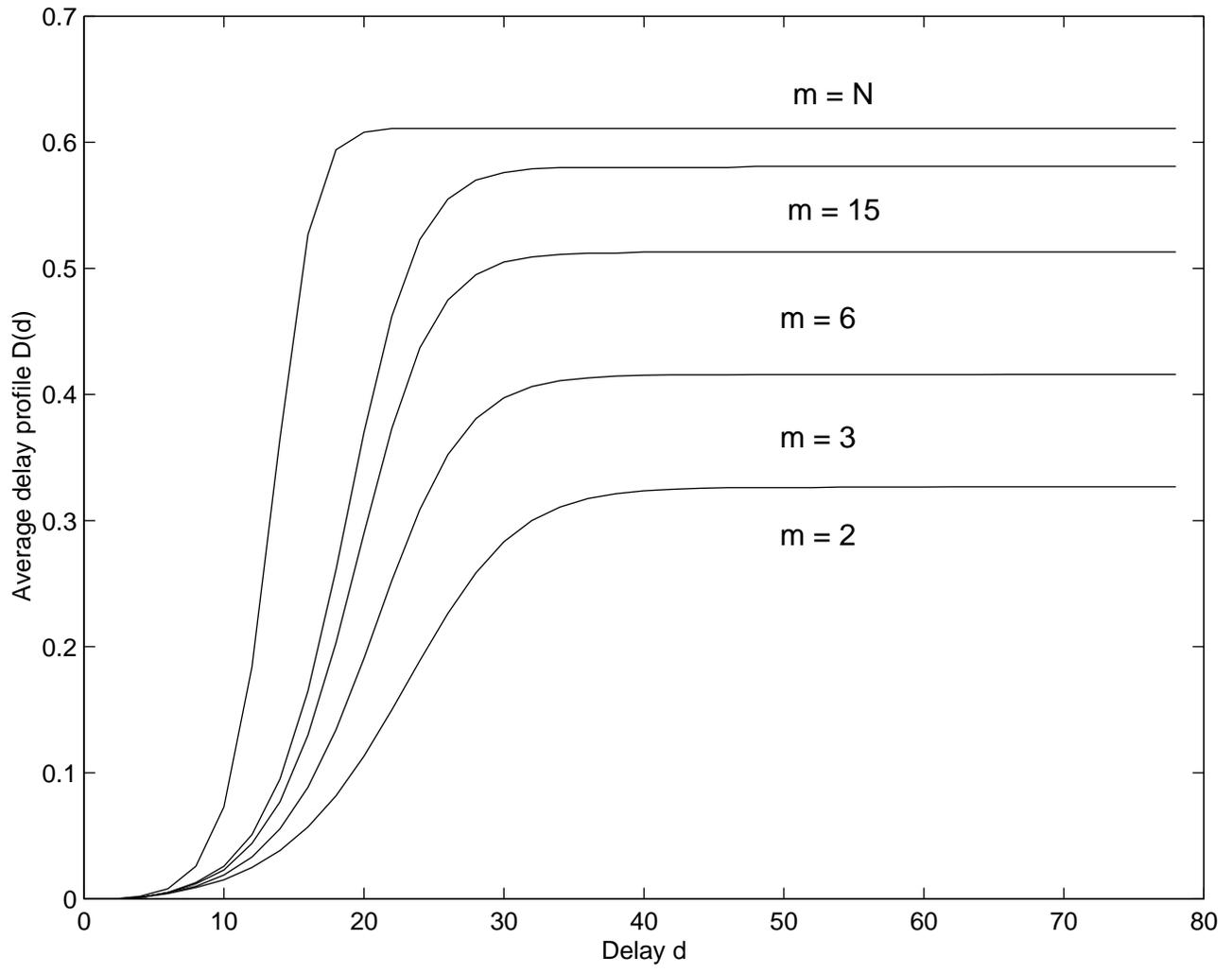